\documentclass[preprint,12pt]{elsarticle}
\usepackage{amsmath,amssymb,amsfonts}
\usepackage[utf8]{inputenc}
\usepackage{babel}
\usepackage{tabularx, multirow}
\usepackage{algorithmic}
\usepackage{textcomp}
\usepackage{xcolor}
\usepackage{fancyhdr,lipsum}
\usepackage[colorlinks=true,]{hyperref}
\usepackage{eurosym}
\usepackage{todonotes}

\graphicspath{{figures/}}

\journal{Applied Energy}
\makeatletter
\def\ps@pprintTitle{%
 \let\@oddhead\@empty
 \let\@evenhead\@empty
 \def\@oddfoot{\centerline{\thepage}}%
 \let\@evenfoot\@oddfoot}
\makeatother

\begin{document}
\emergencystretch 3em

\begin{frontmatter}
    
    \title{Leveraging the Existing German Transmission Grid with Dynamic Line Rating}

    \author[tub]{Philipp Glaum}
    \ead{p.glaum@tu-berlin.de}
    \author[tub]{Fabian Hofmann}
    \ead{m.hofmann@tu-berlin.de}
    \address[tub]{{Technische Universitaet Berlin, Institute of Energy Technology},
        {Berlin},
        {10623},
        {Germany}}

    \begin{abstract}
        % The German government announced that 80\% of the power supply should come from renewable energy sources by 2030. In terms of installations, this means roughly a doubling of today's wind capacity and a tripling of today's solar capacity. For an efficient integration into the system, the fast ramp up of renewables needs to be accompanied by a reorganization of the transmission system.
        The integration of large shares of wind and solar power into the power system benefits from transmission network expansion.
        However, the construction of new power lines requires long planning phases and is often delayed by citizen protests. As an non-invasive alternative, Dynamic Line Rating (DLR) offers the potential to leverage the existing grid by dynamically adjusting the transmission line capacities to the prevailing weather conditions.
        In this study, we present the first investment model  that includes DLR in a large-scale power system with real-world network data and a high temporal resolution.
        Using Germany as an example, we show that a system-wide integration of DLR improves the integration of existing and additional renewables while reducing grid congestion. The evolving synergies between DLR and increased wind generation result in total cost savings of about 3\% of all system costs for a scenario with 80\% renewable power production, mainly due to reduced storage and solar capacity needs. If considering a fully decarbonized electricity system, the cost savings from DLR amount to up to 5.5\% of the system costs, i.e. 4~billion Euro per year. Our results underscore the importance of a rapid implementation of DLR in power systems to support the energy transition and relieve grid congestion.
    \end{abstract}

    \begin{keyword}
        power system modelling, thermal rating, power system planning, renewable energy source, energy transition
    \end{keyword}
\end{frontmatter}
\pagebreak
\section{Introduction}

The efficient integration of renewable energy sources into today's power system requires restructuring key components, in particular the transmission system. With evolving production centers at sites with favorable renewable resources and decentralized power generation on house-hold level, electricity is flowing along new paths that are pushing the transmission network to its physical limits.
To meet high share renewable targets, an expansion of the transmission system is essential for most countries.
However, the process of expanding transmission lines can take 5-10 years and has often faced delays due to administrative issues and protest activities in the past~\cite{michiorriForecastingDynamicLine2015,neukirchProtestsGermanElectricity2016,fernandezReviewDynamicLine2016}. The situation of lacking transmission capacity is therefore expected to worsen in the future which, in the light of rapid installation of renewable generation, will likely increase transmission system congestion, curtailment and grid instability~\cite{erdincComprehensiveOverviewDynamic2020}.
In the example of Germany, most of the wind infrastructure built up throughout the last two decades is located in the North while there is significant industrial demand in the South~\cite{kuhneConflictsNegotiationProcesses2018}.
The current federal government plans to have 100~GW onshore and 30~GW offshore wind power capacity in place by 2030~\cite{federalministryofeconomicaffairsandclimateactionGermanyCurrentClimate2022}, which means roughly a doubling of today's capacities. Like in other countries, there is therefore a strong incentive to make better use of the existing grid infrastructure in near future. \\

To mitigate the lack of transmission capacities, several, complementing measures can be taken:  (1) large scale implementation of storage facilities to flatten power feed-ins and power demands~\cite{caoPreventPromoteGrid2021}; (2) usage of alternative energy carrier networks like hydrogen to relieve the electricity grid~\cite{welderDesignEvaluationHydrogen2019}; (3) leveraging the existing grid infrastructure to increase the transmission line capacity. In the following, we will focus on the last option, which in comparison to (1) and (2) stands out through a fast, low-cost and non-invasive implementation~\cite{deutscheenergie-agenturgmbhdenaErgebnispapierDenaStakeholderprozessHohere, minguezApplicationDigitalElevation2022}.

The transmission of electricity dissipates resistive losses in form of heat. In meshed networks like in Germany, the limiting factor for line capacity is most often the thermal limit, i.e. the maximally allowed temperature of the conductor ensuring a stable transmission.
% Is the thermal limit really a temperature?
This maximal line capacity is also regarded as thermal capacity.
If the transmission line exceeds its thermal capacity, a safe operation of the transmission line is no longer guaranteed due increased line sag and clearance infringements~\cite{fernandezReviewDynamicLine2016}.

Traditionally, the thermal capacity of a transmission line is calculated assuming unfavorable, static weather conditions such as 40$^\circ$C  ambient temperature and 0.6~m/s wind speed~\cite{fernandezReviewDynamicLine2016}. This is referred to as Static Line Rating (SLR).
By design, SLR underestimates the thermal capacity of a transmission line and, when implemented in practice, leads to an underutilization of the transmission infrastructure. On the other hand, Dynamic Line Rating (DLR) calculates the line capacity taking into account prevailing weather conditions. Cold weather and wind cool overheated transmission lines, enabling the thermal capacity to be raised. This results in key benefits in cost-efficiency, congestion reduction and better wind power integration~\cite{erdincComprehensiveOverviewDynamic2020}.
Today, DLR is applied by the German Transmission System Operators (TSOs) in a few projects and in a simplified form, distinguishing between regional winter and summer capacities~\cite{bundesnetzagenturMonitoringReport20212022}. This is set to change according to the Federal Network Agency's network development plan, which calls on TSOs to implement real-time DLR wherever possible~\cite{bundesnetzagenturMonitoringReport20212022}. In particular against the backdrop of the current energy crisis, the importance of DLR was pointed out by the Federal Ministry of Economical Affairs and Climate Action \cite{federalministryofeconomicaffairsandclimateactionMehrStromIns}.

In practice, there are several methods to ensure the thermal capacity for a transmission line is not exceeded~\cite{albizuMethodsIncreasingRating2005}:
\begin{itemize}
    \item \textbf{Weather measurement:} Nearby weather stations measure ambient conditions to calculate the theoretical thermal capacity.
    \item \textbf{Conductor temperature measurement:} Sensors along the line measure the real time temperature of the conductor.
    \item \textbf{Tension measurement:} A load cell measures the tension of the line to derive the clearance of the line.
    \item \textbf{Sag measurement:} The security clearance is directly monitored.
\end{itemize}

Among these, weather based measuring of the thermal capacity is considered the simplest as well as the cheapest method to implement~\cite{KARIMI2018600}.
The assumed investment costs for the implementation of DLR with the different methods vary between 60k and 80k~\euro{} per km with an expected service life of 30 years ~\cite{samweberProjektMeritOrder2016}.

In the literature, the field of energy system modelling has gained increasing importance in recent years. Detailed models of energy systems using real-world input data are used to assess new technologies and/or policies in the context of the energy transition \cite{gea-bermudezRoleSectorCoupling2021,brownSynergiesSectorCoupling2018,howellsOSeMOSYSOpenSource2011,ogunmodedeOptimizingDesignDispatch2021}.

Despite its expected system benefits, none of these models consider DLR as a transmission line capacity enhancement method. Typically, DLR is studied based on artificial IEEE systems with low spatial and temporal resolution. The publications in \cite{numanCoordinatedOperationReconfigurable2021,viaforaDayaheadDispatchOptimization2019,dabbaghjamaneshEffectiveSchedulingReconfigurable2019,chakrapanimanakariMinimizationWindPower2020,liDayAheadSchedulingPower2019} perform an operational optimization with DLR of bus systems with 24--118 nodes and a time-horizon of 24 hours. In particular, the studies in \cite{numanCoordinatedOperationReconfigurable2021} presents a mixed-integer optimization model for evaluating the optimal set of line candidates to be upgraded with DLR. A slight expansion is provided in \cite{jabarnejadOptimalInvestmentPlan2016} where an investment model of a 118~IEEE bus system using a mixed-integer optimization with benders decomposition is presented.

The literature lacks a comprehensive assessment of DLR based on a detailed investment model with (1) high spatial and temporal resolution, (2) long time-horizon, (3) high renewable penetration and (4) real-world input data.
This study addresses this gap. Our study is the first to examine the benefits of DLR considering (1)--(4), all which are necessary to assess the compound system effects of DLR.
%in a for high renewable penetration and high grid resolution, both of which are necessary to assess the compound system effects of DLR.
By focusing on Germany only, we allow for a high quality of the input data and a network representation in its original topology.
The workflow, openly available at \cite{hofmannDynamicLineRating}, can be easily extended to other countries.

The theoretical transmission capacity potential with DLR has already been discussed in \cite{samweberProjektMeritOrder2016} for Germany and in \cite{energynauticsgmbhoko-insitute.v.bird&birdDistributionSystemStudy2016} for Rhineland-Palatinate with high renewable penetration. However, both reports do not include any operational or capacity expansion optimization of the power system.
Differing from our first study~\cite{glaumEnhancingGermanTransmission2022}, the following work has improved methodology and extended analysis that uses updated cost assumptions and broadens the scope of scenarios to the year 2035.

The article is structured as follows.
Section~\ref{sec:methodology} describes the methodology of the DLR implementation (\ref{sec:dlr}) and the underlying power system modelling (\ref{sec:powersystem}). Section~\ref{sec:results} presents the results of the different analyzed scenarios with  limitations outlined in Section~\ref{sec:limitations}. A conclusion is presented in Section~\ref{sec:conclusion}.

\section{Methodology}
\label{sec:methodology}

\subsection{Dynamic Line Rating}
\label{sec:dlr}

The key concept of DLR is based on a dynamic estimation of the maximally allowed electrical current for the conducting material, also referred to as ampacity.
The ampacity is set such that the conductor does not surpass the maximally allowed temperature  after which impermissible sag of the line or hardware damage can be expected~\cite{KARIMI2018600}.
There are two widely used standards to calculate the ampacity of a conductor, namely the IEEE and the CIGRE standard~\cite{ieeeIEEEStd738},~\cite{iglesiasGUIDETHERMALRATING2014}. As the IEEE standard is known to be  more conservative~\cite{simmsComparativeAnalysisDynamic2013}, we choose it for our modeling.
The following outlines the basic concept of DLR according to IEEE. For further details refer to~\cite{ieeeIEEEStandardCalculating2012},~\cite{ieeeIEEEStd738}.
For each conductor, the heat balance equation
\begin{align}
    q_c +q_r = q_s + I^2 \cdot R(T)
    \label{eq:heat_balance}
\end{align}

relates the heat losses on the left hand side to the heat gains on the right hand side. Convective heat loss $q_c$, which represents cooling by ambient air, depends on ambient temperature, wind speed and angle, conductor material and geometry. The radiated heat loss $q_r$ is the net energy lost through black body radiation. Solar heat gain $q_s$, on the other hand, is caused by solar heat radiating  onto the conductor. Finally, the resistive heat gain $I^2 \cdot R(T)$ is given for an electrical current $I$ and temperature-dependent resistance $R(T)$, where $T$ is the temperature of the conductor.  The latter can be approximated by linearly interpolating from reference resistance $R_{ref} = R(T_{ref})$ using a material specific temperature coefficient $\alpha$, i.e.
\begin{align}
    R(T) = R_{ref} \cdot (1 + \alpha \cdot (T - T_{ref}))
\end{align}

Solving \eqref{eq:heat_balance} for the electrical current and setting the temperature to its maximally allowed limit $T_{max}$, yields the ampacity
\begin{align}\label{ampacity}
    I_{max} = \sqrt{\frac{q_c +q_r-q_s}{R(T_{max})}}
\end{align}
which for three-phase electric power transmission operating at voltage level $V$ leads to a maximally allowed, constant power transfer of
\begin{align}
    P_{max} = \sqrt{3} \cdot I_{max} \cdot V
\end{align}

\begin{figure}[!h]
    \centering
    \includegraphics[width=0.65\linewidth]{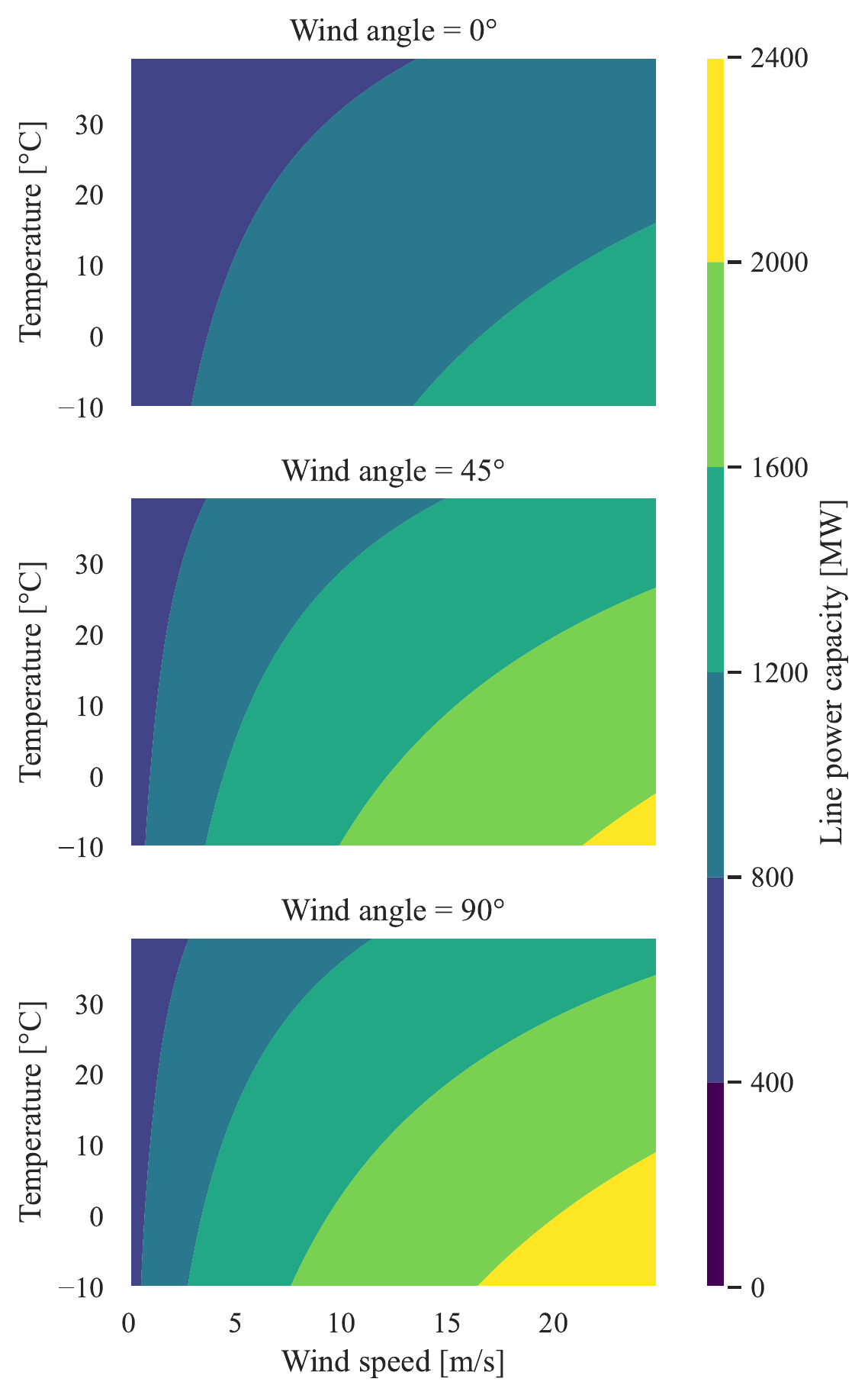}
    \caption{Transmission capacity for different environmental conditions. For three different wind incidence angles (0$^\circ$, 45$^\circ$, 90$^\circ$), the maximal transmission capacity of a typical electricity line ($R(T_{max})$=$9.39\cdot10^{-5}\,\Omega/\text{m}$, $V$~=~380~kV) is shown as a function of temperature and wind speed. The cooling effect of a perpendicular, strong, cold wind can lead to capacity increase of factor 4-5 compared to conservative conditions with 40$^\circ$C and low wind.}
    \label{fig:parameter-space-reduced}
\end{figure}

Fig.~\ref{fig:parameter-space-reduced} shows $P_{max}$ for a single wire of a typical 3-phase transmission line with $R(T_{max}=80^\circ C) = 9.39\cdot10^{-5}\,\Omega/\text{m}$ and $V$~=~380~kV, as a function of temperature and wind speed for three different wind incidence angles, 0$^\circ$, 45$^\circ$ and 90$^\circ$. The cooling effect at cold temperature with strong perpendicular wind leads to a transmission capacity increase of factor 4-5 compared to conservative conditions with 40$^\circ$C and low wind.

For this study the IEEE standard was implemented in Atlite~\cite{hofmannAtliteLightweightPython2021}, a Python package used for converting weather data into renewable power potentials. Atlite obtains the weather data from the ECMWF Reanalysis v5 (ERA5) dataset providing various weather-related variables in a hourly resolution on a $0.25^\circ \times 0.25^\circ$ grid.

\begin{figure}[!ht]
    \centering
    \includegraphics[width=.7\linewidth]{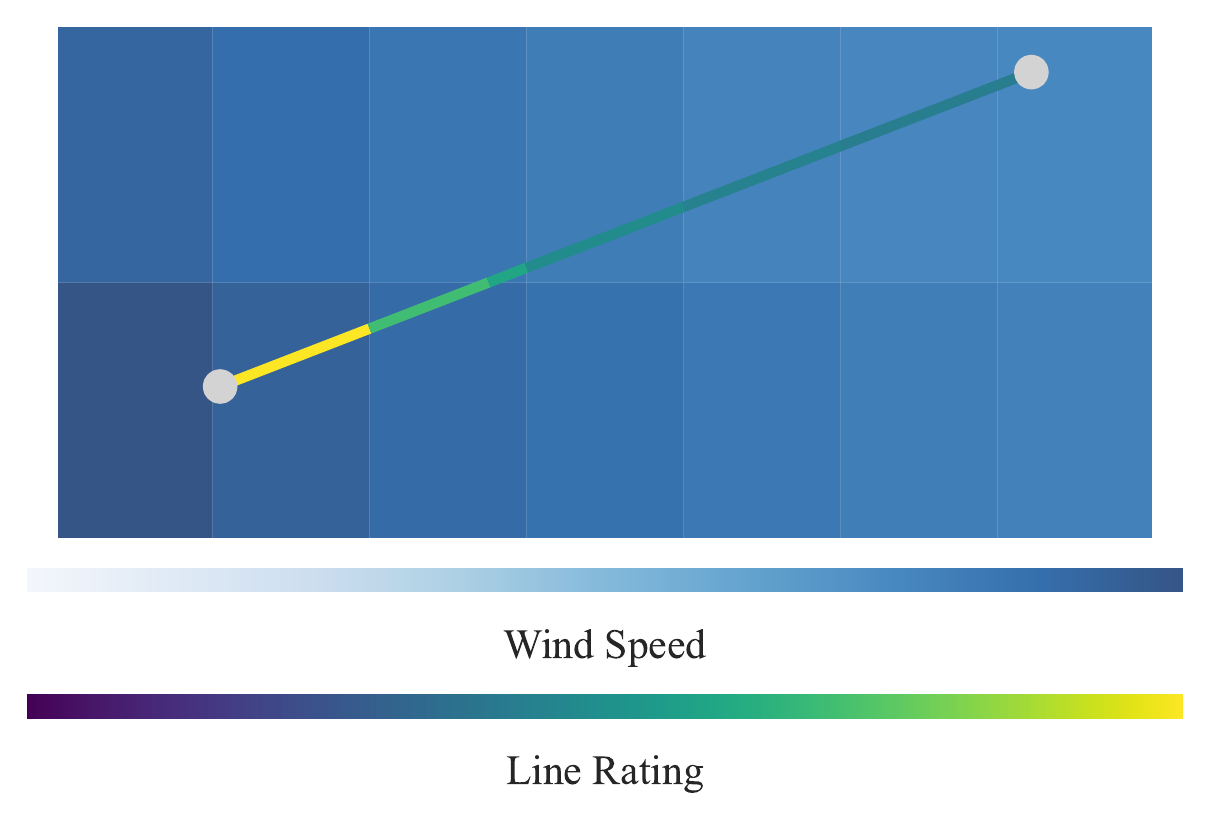}
    \caption{The figure shows a transmission line intersecting different weather cells with individual wind speeds. The implementation in Atlite calculates different ampacities for each of the overlapping line segments, of which the minimum value determines the ampacity for the entire line.}
    \label{fig:line-rating-calculation}
\end{figure}

The transmission lines are mapped onto the weather grid and their transmission capacity is calculated according to Equation~\ref{ampacity}. If multiple weather grid cells overlap with a transmission line, the grid cell with the most unfavorable condition provides the ampacity for the entire line. This is illustrated in Fig.~\ref{fig:line-rating-calculation}.

\subsection{Power System Modeling}
\label{sec:powersystem}
Using the Python package \textit{Python for Power System Analysis} (PyPSA)~\cite{brownPyPSAPythonPower2018,browntomPyPSAPythonPower2022}, the linear ESOM is represented by a set of buses interconnected via transmission lines and complemented by loads, generators and storage facilities. Further, time-dependent electric demand per bus and generation potentials per generator are included for one year with hourly resolution. The generator dispatch and deployment is determined by minimizing the total system cost. The optimization uses the linearized power flow approximation~\cite{purchalaUsefulnessDCPower2005,brownLinearOptimalPower2016} and is subject to systemic constraints (CO$_2$ budget, limited capacity expansion, etc.).

% The underlying model is created from the PyPSA-Eur framework [16], and for Germany consists of 256 representative nodes and considers a time horizon of one year with an hourly resolution. To model the effect of DLR on the power system, we use the IEEE standard~\cite{IEEEStandardCalculating2012} with highly resolved historical weather data implemented in the Atlite software~\cite{hofmannAtliteLightweightPython2021}. The advantages of the DLR are illustrated by comparison with an SLR-based system. Therefore, we analyze and compare curtailment, grid congestion as well as the optimal generator deployment.

Data on the transmission grid, power plants, renewable potentials and demand for the model are created with the workflow \textit{PyPSA-EUR}~\cite{horschPyPSAEurOpenOptimisation2018,horschjonasPyPSAEurOpenOptimisation2022}.
The representative network of Germany comprises all 256~substations and 333~transmission lines operating at 220~kV and above. Electrical parameters of transmission lines are derived by mapping voltage level of the lines to standard line types given in~\cite{oedingElektrischeKraftwerkeUnd2011}.
The demand data and renewable profile data, provided by Atlite, have an hourly resolution.

For the DLR model, the transmission capacity per line and time step is given by the formulation in Section~\ref{sec:dlr}, for the SLR model by the standard static transmission capacities. As DLR is calculated with averaged hourly wind speed data, sub-hourly wind speed fluctuations are neglected. This leads to a slightly overestimated $P_{max}$, which we consequently scale down by an empirically derived factor of $0.95$, see \ref{dlr_factor} for details. Furthermore, we account for N--1 network security in both SLR and DLR models by restricting the power transmission P per line to $-0.7\,P_{max} \le P \le 0.7\,P_{max}$ leaving a 30\% capacity buffer~\cite{brownOptimisingEuropeanTransmission2016}.

As already shown in Fig.~\ref{fig:parameter-space-reduced}, DLR can easily lead to a capacity increase of more than 100\% if the circumstances are beneficial. To ensure network security, TSO's typically limit the relative DLR capacity to 150\%~\cite{transnetbwFreileitungsmonitoring} of the SLR capacity. On the other hand, the ENTSO-E reports manageable capacity increases of 100\%~\cite{entso-eDynamicLineRating}. To account for this uncertainty, we calculate each limit separately for $P_{max}$ at 130\%, 150\%, 180\%, and 200\%. If not mentioned further, no $P_{max}$ limit is considered.

\section{Results}
\label{sec:results}

The presented work includes three scenarios, which are listed in Table~\ref{tab:scenarios}. In the first scenario, the operation of the existing network is optimized for 2019 to validate the model and demonstrate the near-term system benefits of DLR. In the second scenario, we run an investment optimization for the year 2030 to show the long-term benefits of DLR.
By enforcing an 80\% renewable production share, denoted as renewable share, we are able to assess DLR as a supporting measure to reach the German government's target of 80\% renewable power supply by 2030~\cite{federalministryofeconomicaffairsandclimateactionGermanyCurrentClimate2022}. Beyond that, the German government targets climate neutrality of the power sector by 2035. In our third scenario, we therefore alter the renewable share from 85--100\% in steps of 5\%. The parameters settings for the study cases can be found in the \ref{tab:cases_parameters}.

\begin{table}[!h]
    \begin{tabular}{lll}
        Scenario & Type of optimisation & Minimal renewable share\\
        \hline
       2019 & Operational & -- \\
       2030 & Operational \& capacity expansion & 80\% \\
       2035 & Operational \& capacity expansion & 85\%, 90\%, 95\%, 100\% \\
       \hline
    \end{tabular}
    \caption{Table showing the three regarded scenarios with their type of optimization and their renewable shares.}
    \label{tab:scenarios}
\end{table}

\subsection{Optimal Operation for 2019}
For quantifying the system benefits of DLR in the existing grid, we choose to model the pre-pandemic year 2019 when operations were relatively average. Electrical load and renewable potentials are derived from historical data of the year 2019~\cite{muehlenpfordtTimeSeries2020}, transmission and generation infrastructure are aligned to the state of 2019, using selected data from~\cite{hofmannFRESNAPowerplantmatchingV02022,openpowersystemdataDataPackageNational2017,wiegmansGridkitExtractEntsoE2016}. To approximate the operation of nuclear and lignite power plants mostly running at base load, we enforce a minimal requirement of operation, based on historical data from the ENTSO-E Transparency Platform ~\cite{entso-eENTSOETransparencyPlatform2020}.

\begin{figure}[h!]
    \centering
    \includegraphics[width=.7\linewidth]{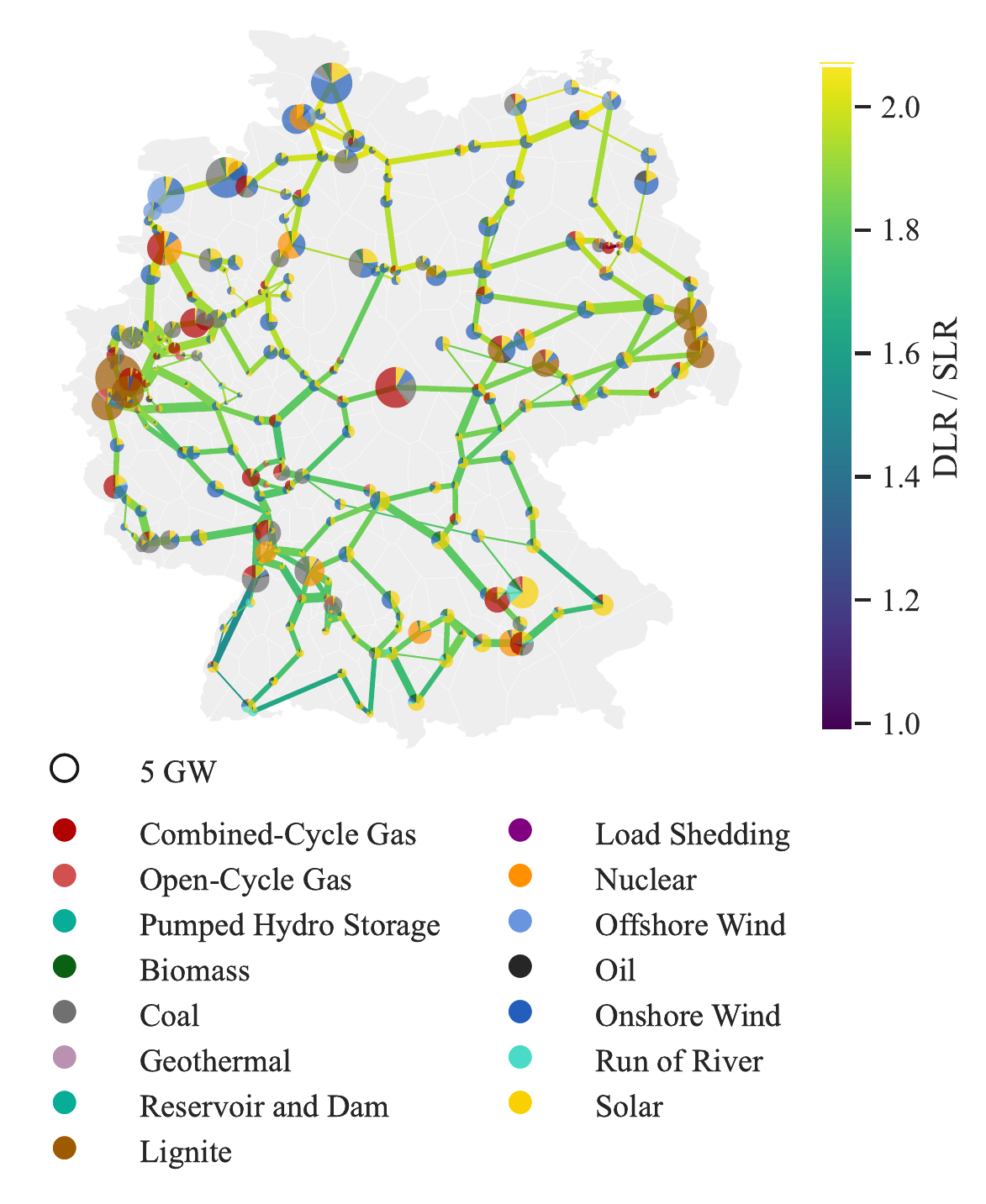}
    \caption{Capacity layout of the 2019 scenario of the German power system. The circle with its subdivisions are proportionally to the installed generation capacity. The widths of the line indicate their nominal transmission capacity, their color the relative average change when going from SLR to DLR.}
    \label{fig:capacity-map-2019}
\end{figure}

Fig.~\ref{fig:capacity-map-2019} shows the geographical layout of existing generation and transmission capacities for the 2019 scenario. The circles with its subdivisions show the installed power generation capacity per site and technology with their area being proportional to the capacity. The widths of the transmission lines indicate the installed nominal capacities, while their colors show the relative average change in capacity when applying DLR. From the latter, we see that the changes in average line capacity vary across Germany. In the north, line capacity increases by 60-90\%, while in the south it increases by 30-50\%. Note that this effect is related to higher average wind speeds in northern Germany. \\

\begin{figure}[h!]
    \centering
    \includegraphics[width=.7\linewidth]{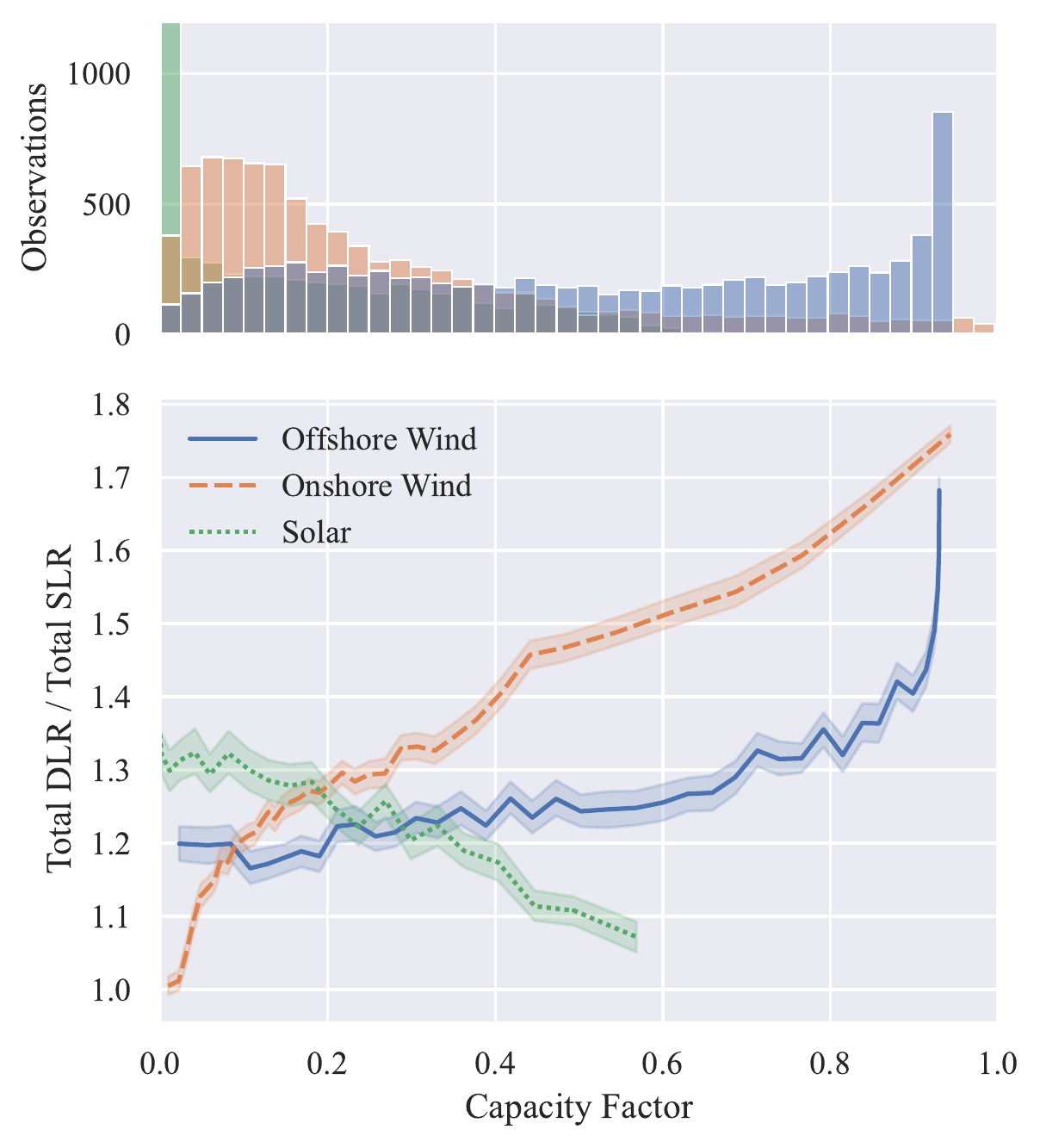}
    \caption{Lower graph shows the ratio between total DLR and SLR transmission capacity as a function of the capacity factor. Onshore and offshore wind production correlate positively with the increase in transmission capacity, solar productions reveal a negative correlation. The upper graph shows the number of observations for a given DLR to SLR ratio and the capacity factor.}
    \label{fig:potential-correlation-2019}
\end{figure}

In addition to location, of particular interest is the time periods over which transmission improvements occur. The upper graph in Fig.~\ref{fig:potential-correlation-2019} shows the number of observations of a given capacity factor, i.e. the total power output relative to the installed capacity, for solar, onshore and offshore wind, when collecting into 40 bins of equal width. For better visualization the y-axis was clipped at 1200 observations. The lower graph shows the average total DLR capacity relative to SLR for the bins as a function of the capacity factor. The shaded areas around the lines correspond to the 95\% confidence intervals.

For onshore wind power, the transmission capacity increase correlates positively and more or less evenly with the capacity factor. During periods of almost no onshore wind production, the average transmission capacity is close to the SLR transmission capacity. The DLR transmission capacity then rises sharply up to a capacity factor of 0.2. From there, the line rating increases almost linearly finalizing at an overall transmission capacity increase of 75\% for periods onshore wind capacity factors close to 1. This correlation reflects the dependency of DLR on the wind speed. As can be seen in the upper graph, most time steps have an onshore wind capacity factor of $\le$0.5. Higher capacity factors only occur in 16\% of the time steps.

For offshore wind power, the correlation between transmission capacity and power production is also positive, but not as steady as for onshore wind power. During periods with low offshore wind production, the transmission capacity increases by roughly 20\%. The transmission improvement rises when going to periods with more offshore wind power. After a steep positive trend when reaching periods of highest offshore wind supply, an increase in transmission capacity of almost 70\% is observed. The overall distribution of capacity factors is relatively flat and peaks at the maximally allowed capacity factor 0.93, set to account for wake losses~\cite{boschTemporallyExplicitSpatially2018}.

For solar power, the trend is roughly the opposite. During periods of low solar supply, the increase in transmission capacity averages 20\%, while it decreases during periods of higher supply. However, the DLR transmission capacity stays above the transmission capacity of SLR. The overall negative correlation results from the fact that on sunny days the cooling effect of temperature and wind tends to be lower. The radiating heating from solar influx plays a marginal role only. In the upper graph, the number of observations was clipped at 1200, leaving out the data point of 4400 time steps without solar supply.

\begin{figure}[h!]
    \centering
    \includegraphics[width=.7\linewidth]{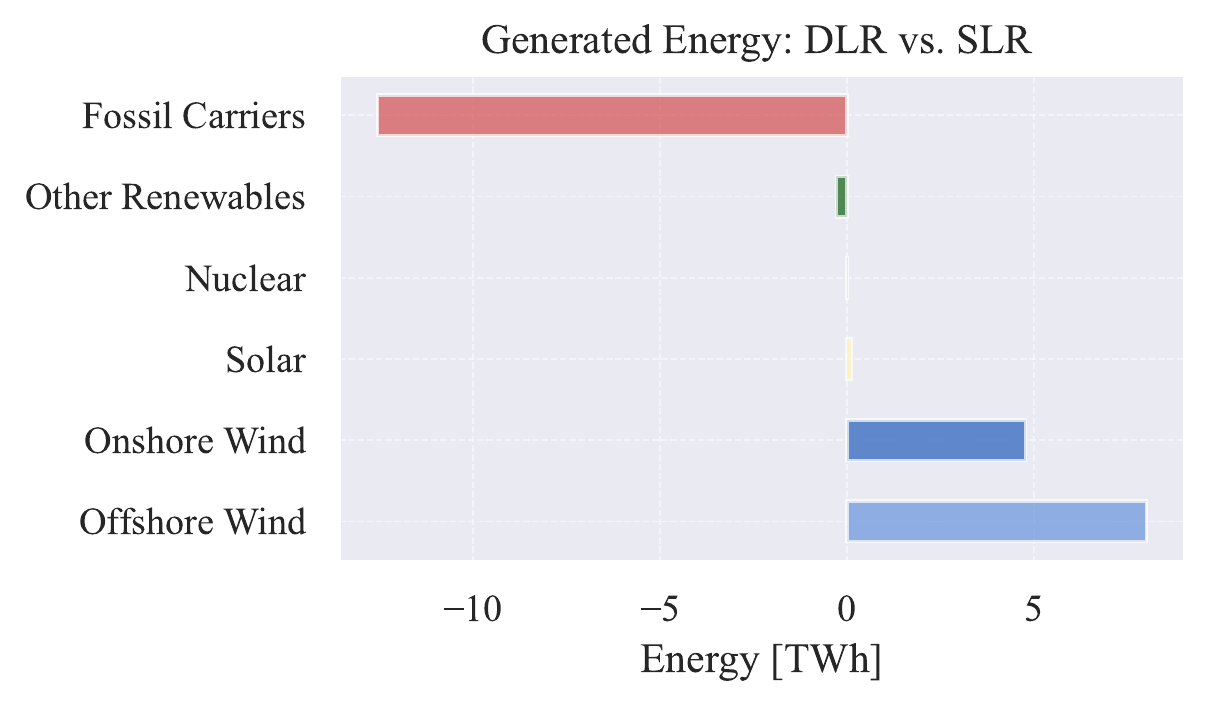}
    \caption{Difference in supply between optimized operation with DLR and SLR in the 2019 scenario. }
    \label{fig:operation-difference-2019}
\end{figure}

In the operational optimization, these effects prove themselves to be quite impactful. Fig. \ref{fig:operation-difference-2019} shows the difference in supply when going from SLR to DLR in the 2019 scenario. Of the 490~TWh total net electricity generation throughout the year, 31\% (149~TWh) is provided by fossil power generation (hard coal, lignite and gas) with SLR. This share drops to 28.9\% (137~TWh) when implementing DLR, while the share of onshore wind power increases from 24\% (116~TWh) to 25\% (122~TWh) and of offshore wind power from 4.4\% (21.4~TWh) to 5.8\% (28.3~TWh). The total generation of the other carriers change marginally. We recall that the installed generation capacities are the same for both scenarios.
Solely, from the change in operation, the system saves around 583~m\euro{}/yr operational expenditure (OPEX) which translates to roughly 7~\% of total cost.
Note here that the assumed gas price is 27\euro{}/MWh. Considering increased gas prices (roughly 150\euro{}/MWh) due to current energy crisis would lead to significantly higher cost savings.
In addition to the OPEX savings, the implementation of DLR leads to a carbon emission reduction of 10.8~$\text{MtCO}_2$ (6.5\%) compared to the SLR model. \\

\begin{figure}[h!]
    \centering
    \includegraphics[width=.7\linewidth]{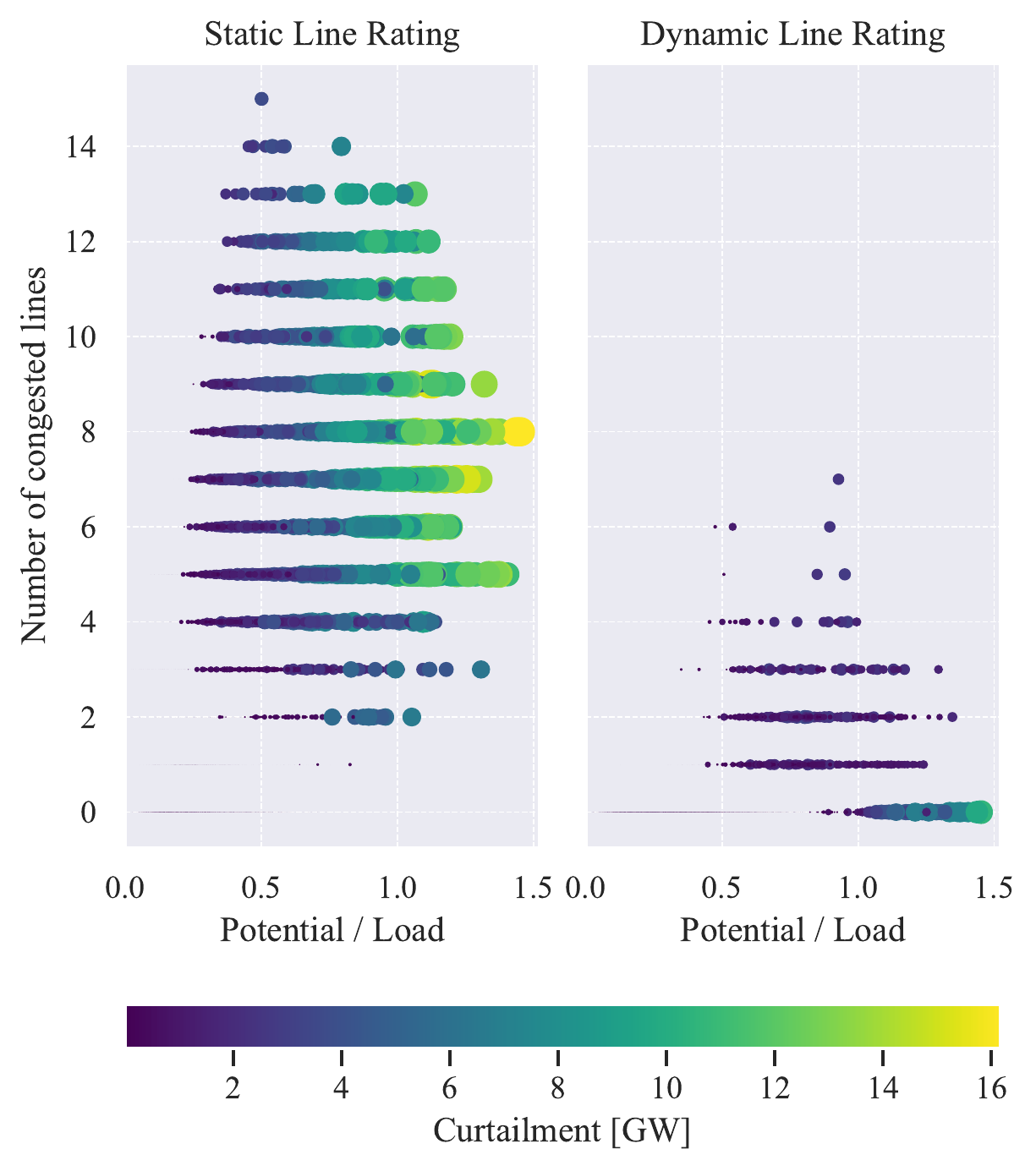}
    \caption{Number of congested lines for SLR and DLR as a function of the ratio between the renewable generation potential and the load for each hour during the year. A ratio higher than 1 implies an over-supply which is curtailed. The size and color of the scatter points correspond to the amount of curtailment at the regarded hour.}
    \label{fig:congestion-correlation-2019}
\end{figure}

The main reason for the shift in generation is the improved transmission capacity during periods when wind power was previously curtailed. Fig.~\ref{fig:congestion-correlation-2019} shows the number of congested lines as a function of the total renewable power potential relative to the total load in each time step. The color and size of the dots indicate the renewable curtailment. It stands out that DLR (right panel) leads to a significant decrease in transmission congestion as well as curtailment. While the SLR model (left panel) reveals up to 16 congested lines, maximally 8 lines are congested in the DLR model. In particular, at times with a shortage of renewable energy (Potential / Load $<$ 1) the SLR system has to curtail 7 times more power due to transmission congestion. This does not hold for the DLR model. The same accounts for times with renewable excess power (Potential / Load $>$ 1), where power is mainly curtailed due to oversupply in the DLR model and curtailed due to oversupply and congestion in the SLR model. \\

\subsection{Optimal Investment 2030}\label{sec:2030}

In the scenario for 2030 the capacity of renewable power plants, gas turbines, batteries and hydrogen infrastructure are optimized. From the existing power plant fleet, only gas power plants with a decommissioning date later than 2030 are included. To align our model with the government's coal-phase out plans~\cite{federalministryofeconomicaffairsandclimateactionGermanyCurrentClimate2022}, we do not consider the use of existing coal and lignite assets. The grid infrastructure is supplemented by grid expansion projects from the TYNDP~\cite{entsoeTYNDP2018Project2018} to be built by 2030.
The electrical load time series, originally representing the default year 2013 of PyPSA-Eur~\cite{horschjonasPyPSAEurOpenOptimisation2022}, is scaled up to meet the predicted total net demand for 2030 of 555~TWh~\cite{kemmlerEntwicklungBruttostromverbrauchsBis2021}. The supply from renewables, i.e. solar, wind, biomass and hydro facilities, is enforced to meet the 80\% target.
The cost parameters for 2030 are retrieved from the technology data-base published at~\cite{lisazeyenPyPSATechnologydataTechnology2022}. In addition, to properly reflect the future operational costs of fossil generators, we impose a price of 120~\euro{} per tonne CO$_2$.

\begin{table}[h]
    \caption{Comparison of a selection of capacity totals for today, reported at~\cite{burgerInstalledPowerEnergyCharts} by the time of writing, and the optimized 2030 scenario with a 80\% renewable share with Static Line Rating and Dynamic Line Rating. }
    \centering
    \begin{tabular}{lrrrrc}
        & Today & SLR & DLR & $\Delta$ & Unit \\
        \hline
       Solar & 64 & 203 & 179.9 & --23 & GW \\
       Onshore Wind & 57.7 & 102.1 & 96.4 & --5.7 & GW \\
       Offshore Wind & 8 & 9.4 & 13.7 & 4.3 & GW \\
       Natural Gas & 32 & 45.9 & 47.6 & 1.7 & GW \\
       Battery Discharge & - & 20.9 & 14.2 & --6.7 & GW \\
       Battery Storage & - & 133.8 & 88.1 & --45.7 & GWh \\
       \hline
    \end{tabular}
    \label{tab:installed capacity}
\end{table}

Table \ref{tab:installed capacity} shows the optimal capacities of technologies varying across the models, in comparison to today's capacities. %The full set of capacity totals as well as the spatial capacity layout can be found in \ref{sec:2030-scenario-plots}. %TODO: add figures and tables t appendix
Roughly speaking, the optimal solution to meet the 2030 target requires tripling solar capacity, doubling wind capacity, and expanding batteries. While solar and onshore wind capacity totals are roughly in alignment the government's target for 2030  (200~GW solar, 100~GW onshore wind), the model expands much less offshore wind power then planned by the government (30~GW)~\cite{federalministryofeconomicaffairsandclimateactionGermanyCurrentClimate2022}. In both scenarios, no hydrogen facilities are installed, which, as we show later, turn out to be profitable for renewable shares higher than 90\%.

On the one hand, the DLR model results in 4.3~GW additional offshore wind expansion and 1.7~GW natural gas expansion. On the other hand, other technologies are expanded significantly less: solar by 23~GW, onshore wind by 5.7~GW, and most importantly battery by a total storage capacity of 45~GWh. This equals to a third of total battery installations in the SLR model.

\begin{figure}[!h]
    \centering
    \includegraphics[width=\linewidth]{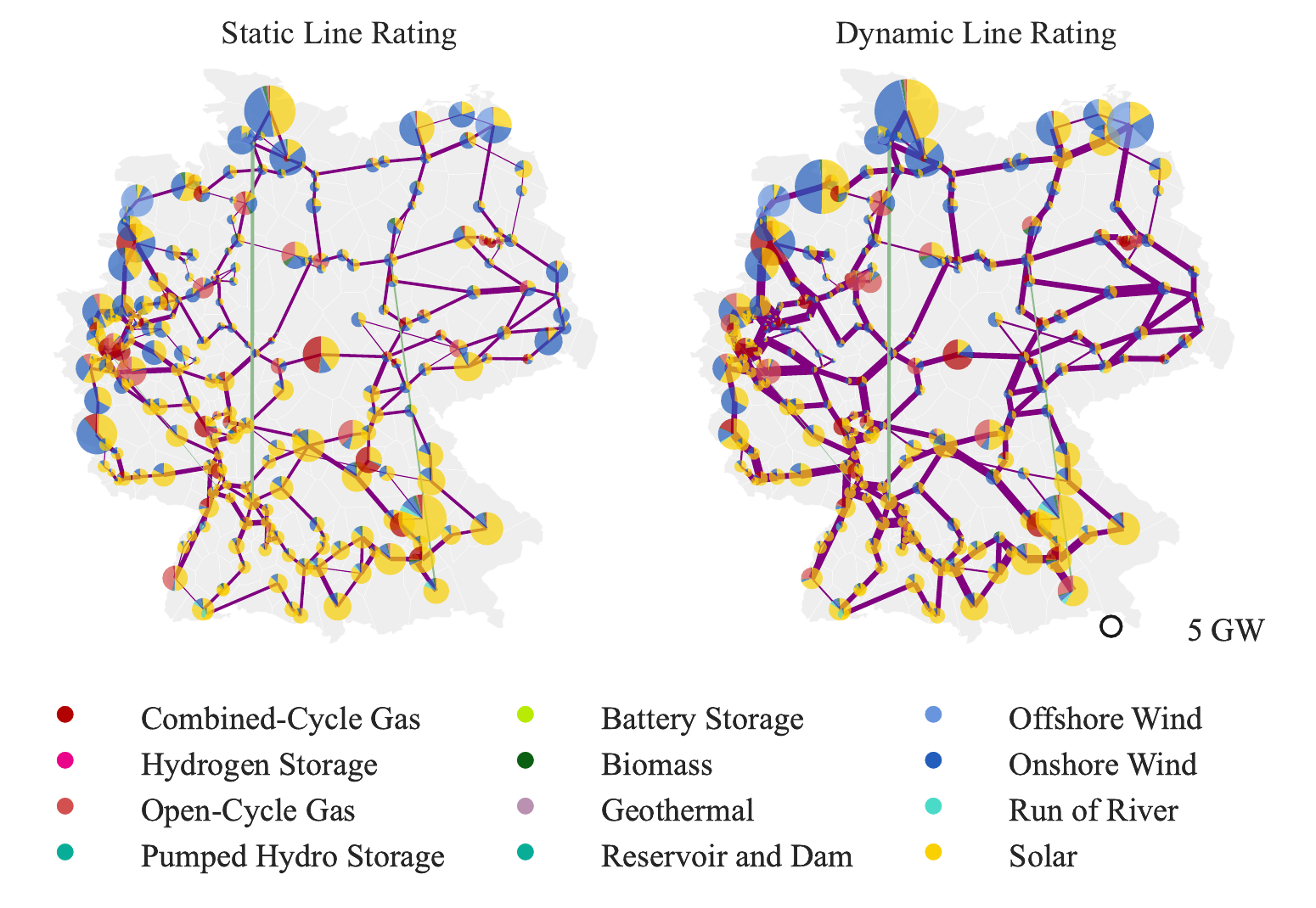}
    \caption{Optimized capacities for the 2030 scenario with 80\% renewable share.}
    \label{fig:capacity-map-2030}
\end{figure}

Fig. \ref{fig:capacity-map-2030} shows the optimal spatial distribution of generation and transmission capacities in the SLR and DLR model.
In the DLR model, the line capacities are higher as well as more renewable capacities are built in the north of Germany.

\begin{figure}[!h]
    \centering
    \includegraphics[width=.7\linewidth]{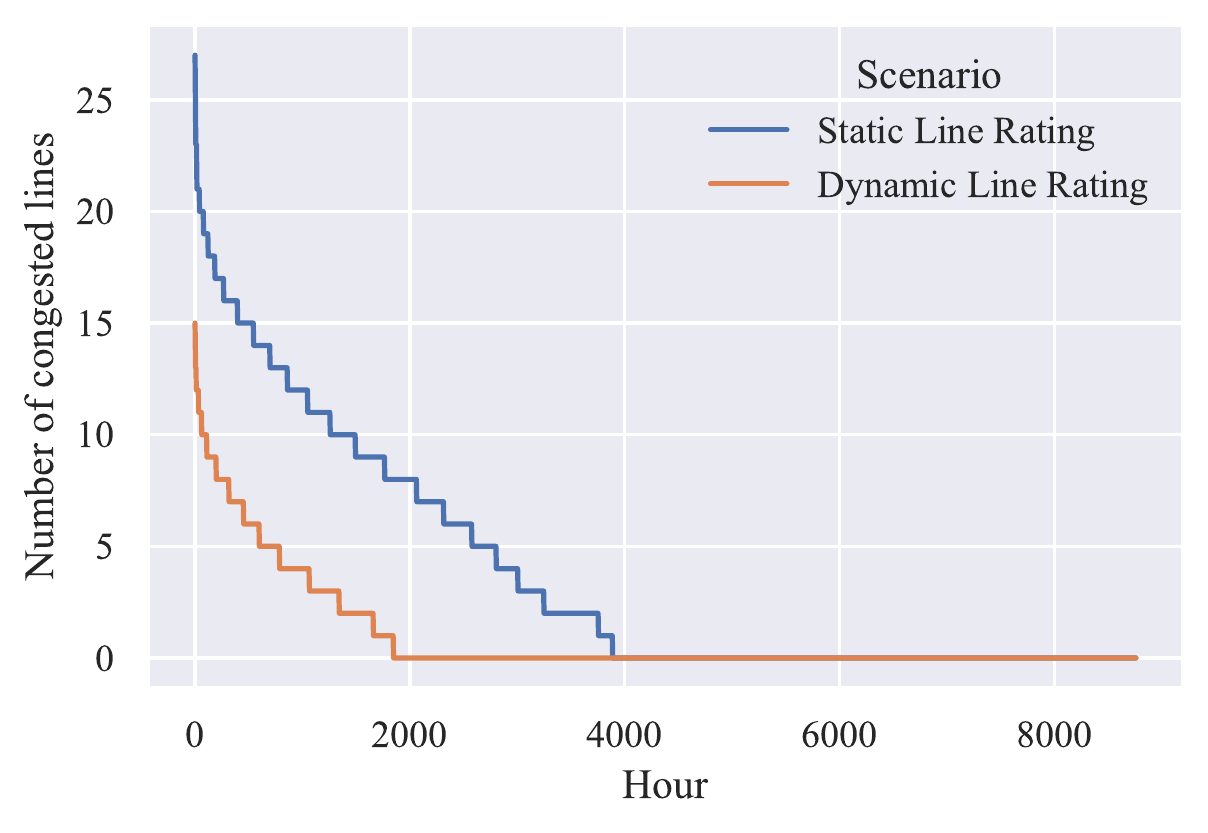}
    \caption{Duration curve for number of congested lines in the 2030 study case.}
    \label{fig:load-duration-2030}
\end{figure}

Together with the changes in the capacity layout, we perceive significant changes in the operation. When implementing DLR, the capacity factor of onshore wind power is lifted from 21\% to 22.5\% leading to additional supply of 1.5~TWh by onshore wind despite its lower capacity. The capacity factor of offshore wind power increases from 39.1\% to 41.5\% leading to an additional supply of 17.7~TWh. In contrast, roughly 17.5~TWh less are produced from solar facilities. By reducing the use of storage technologies and the associated losses, the system produces 91~GWh less in total with DLR. To this end, Fig.~\ref{fig:load-duration-2030} shows the duration curve of the total congestion throughout the simulation year. We see that through DLR the transmission congestion is significantly reduced. More than three quarters of the year are free from congestion while, for SLR, congestion appears in nearly 4000 hours. Furthermore, the number of maximally congested lines drops from 58 to 14 when going from SLR to DLR.

From these numbers we conclude the following. The implementation of DLR enables the model to better integrate offshore wind power in the system. By leveraging offshore wind potentials, the DLR model profits from relatively steady power feed-in. Improved transmission capacity allows offshore wind to penetrate deeper into the system, significantly reducing the need for solar and short-term storage, as well as onshore wind. Overall, the DLR model is 1.15 bn\euro{}/yr cheaper than the SLR model. Most of the cost savings are due to the lower solar and battery installations.

\begin{figure}[!h]
    \centering
    \includegraphics[width=.7\linewidth]{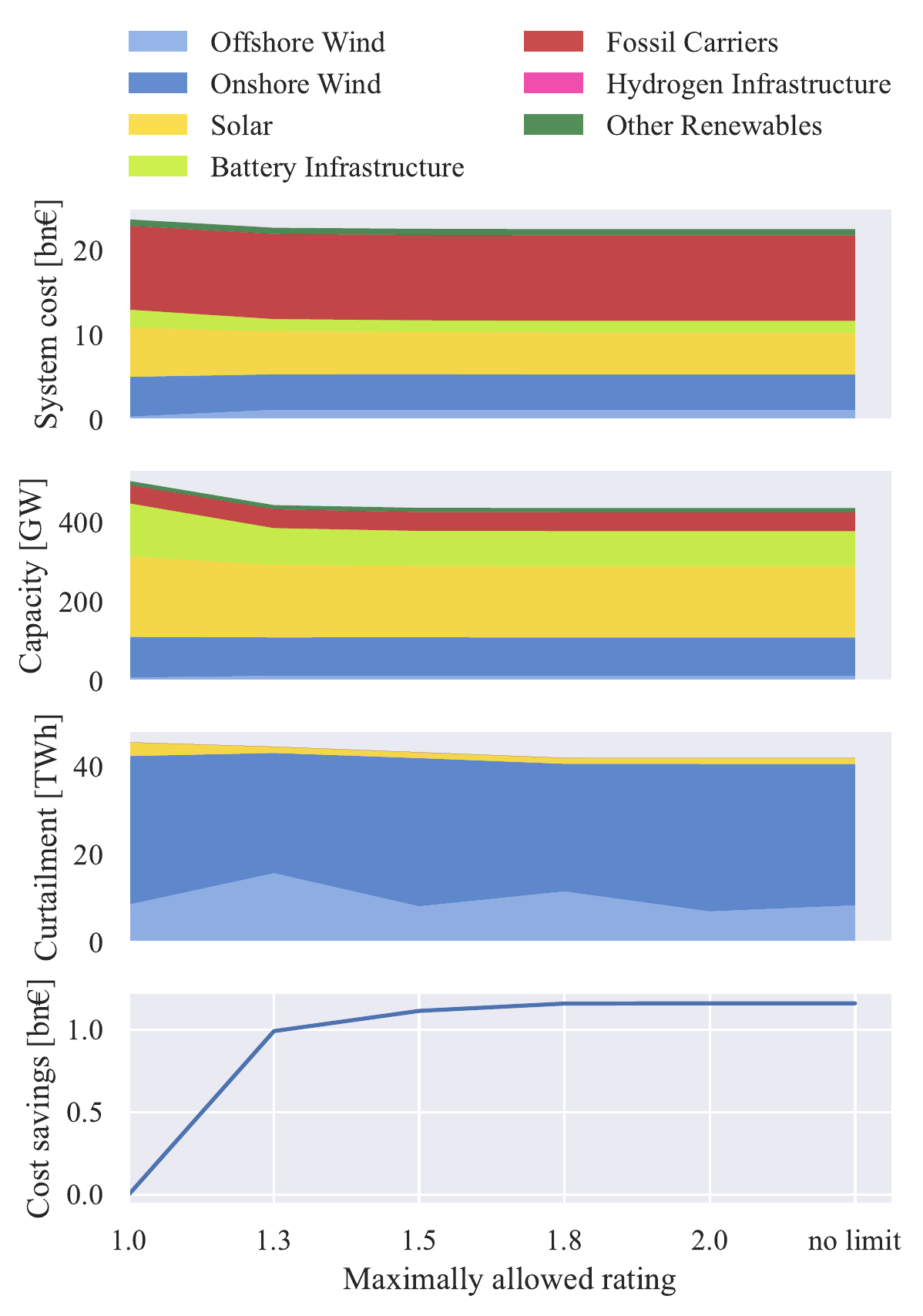}
    \caption{Total cost, optimal capacity, curtailment, and relative total cost savings compared to the SLR model as a function of the maximum line rating allowed per line in the DLR model. Already with a conservative cap at 130\% of the SLR capacity, strong system benefits are achieved.}
    \label{fig:sensitivity_dlr_combined}
\end{figure}

In most of today's real-world implementations, TSO's set a static upper limit to the DLR capacity, e.g. for network security reasons. In the following, we show that the positive effects DLR are robust against such a limit. Fig.~\ref{fig:sensitivity_dlr_combined} shows the system cost, optimal generation capacity, total curtailment, and relative total cost savings as a function of the maximum allowed DLR transmission capacity per line. It can be seen that the largest benefits occurs in the first step going from SLR (1.0) to DLR with a rather conservative upper limit of 130\% of the SLR capacity. From then onwards, optimal generation capacity and curtailment hardly change and system costs are only slightly reduced. In the light of Fig.~\ref{fig:potential-correlation-2019}, these findings can be attributed to two facts: First, only a minority of time steps ($<$~20\%) reveal a relative DLR transmission capacity increase of more than 50\%. Second, even when higher transmission rates are possible, the overall wind power production is so high that the system does not profit from it. With a capacity factor of above 0.5, wind power produced more than 65~GW which is more than the average load of 63.3~GW of the system. Therefore, in most of these times the excess energy is curtailed independently of the DLR transmission capacity limit.

\subsection{Optimal Investment 2035}
Looking beyond the scope of 2030, we vary the production share of renewable technologies gradually in the SLR and DLR model from 80 to 100\%. This variation represents the pathway from 2030 to 2035 in which the German government targets an almost decarbonized power sector~\cite{federalministryofeconomicaffairsandclimateactionUeberblickspapierOsterpaket2022}.
For this scenario, no DLR limits are considered. Note that all the other  parameters, including the electrical load and weather year, remain the same as in the 2030 scenario to ensure comparability.

\begin{figure}[!h]
    \centering
    \includegraphics[width=\linewidth]{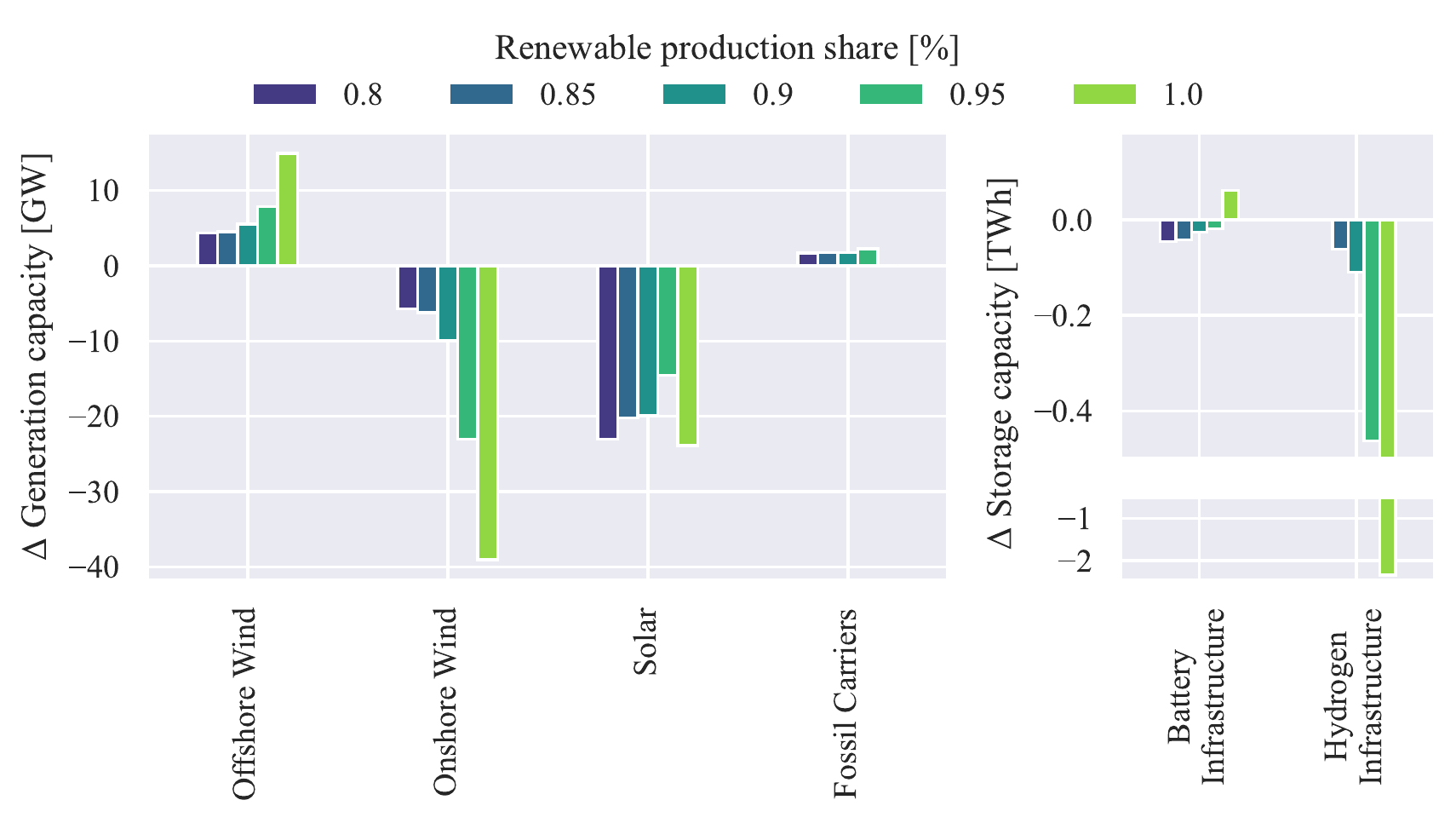}
    \caption{Change in expanded generation and storage capacity for different shares of renewable production.}
    \label{fig:capacity-2035}
\end{figure}

Fig.~\ref{fig:capacity-2035} shows the capacity changes between the SLR and DLR model for all considered renewable shares. Other renewables are not depicted as their capacities are equal for both models.
Continuing the trend of the 2030 scenario, we find that the DLR models require less generation capacity for all technologies except offshore wind. The optimal offshore wind capacity in the DLR model increases continuously with the share of renewables, from +4.3~GW at a share of 80\% to +14.9~GW at a share of 100\%.
For onshore wind, we perceive an opposite trend. With increasing renewable share, DLR builds progressively less capacity resulting in a 39~GW reduction.
The difference in solar capacities for the DLR first decreases from --23~GW at 80\% to --14~GW at 95\% compared to the SLR model. Here, the DLR model sees an advantage in integrating more solar into the grid. However, with the strong increase of offshore wind power in the DLR model at a share of 100\% renewables, less solar capacity is needed again with --23.8~GW.
Regarding fossil generators, DLR builds slightly more capacity with +1.68~GW at 80\% to +2.2~GW at 95\% production share.
However, even though more capacities are built in the 95\% case, the fossil generators of the DLR model generate --40~GWh less electricity.
This indicates that the DLR model only builds more fossil generator capacities because it requires a higher fossil power output for single hours with scarce wind resources.
%add info that this could be solved with cross border flows
For the 100\% renewable share case, in both SLR and DLR model the fossil generation capacity drops to zero due to the full decarbonization of the model without considering carbon capture technologies.

At this point, we may conclude that with increasing share of renewables, the DLR implementation allows trading a relatively small increase in offshore wind power against a high decrease of onshore wind and solar PV capacity.
When looking at the storage infrastructure, we see that in general DLR needs less storage capacity than SLR. For battery infrastructure, DLR builds --45~GWh less capacity at 80\% share. However, with increasing renewable share DLR gradually builds more battery capacity which finally leads to a battery capacity increase of +61~GWh at 100\% share.
As for the hydrogen infrastructure, DLR requires a smaller expansion for all renewable energy shares above 80\%, with the difference increasing as the share increases, to --2.35~TWh storage capacity at 100\% renewable share.

\begin{figure}[!h]
    \centering
    \includegraphics[width=0.7\linewidth]{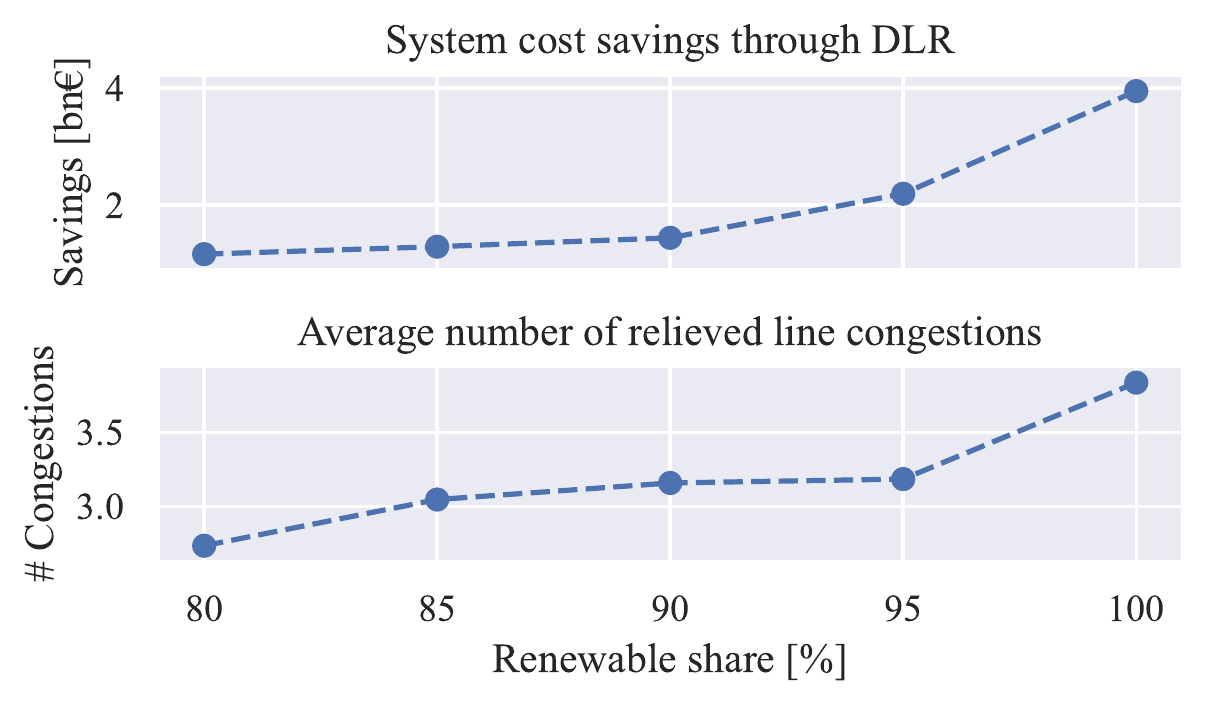}
    \caption{The top figure shows the cost change from SLR to DLR for different shares of renewable production. The bottom figure depicts the change in averaged congested lines per hour.}
    \label{fig:cost-curtailment-sensitivity-2035}
\end{figure}

In Fig. \ref{fig:cost-curtailment-sensitivity-2035}, the cost and grid congestion difference is depicted when going from 80 to 100\% renewable share.
The cost savings from DLR compared to SLR increase from 1.2~bn\euro{}/yr at 80\% renewable share to 3.9~bn\euro{}/yr at 100\% renewable share, underlying benefits of DLR with higher degree of decarbonization.
In terms of numbers of congested lines, DLR similarly shows an increasing benefit in congestion relief with higher renewable shares.
Here the relation between renewable share and relieved line congestions is not as steady. From 80 to 95\%  the benefit of DLR compared to SLR on line congestion relief decrease starting at 2.7 and going up to 3.1 relieved line congestions at 80 and 95\%.
However, from 95 to 100\% renewable share there is a clear increase in congestion relief, with 3.8 relieved lines for DLR compared to SLR at 100\% renewable share.
One explanation for this is the missing fossil generation fleet which could still be used at 95\% renewable share to circumvent transmission bottlenecks.

\begin{table}[!h]
    \caption{Comparison of a selection of total capacity planned by the German government in 2035 \cite{federalministryofeconomicaffairsandclimateactionRegulationDraftEEG2022, federalministryofeconomicaffairsandclimateactionUeberblickspapierOsterpaket2022} and from the optimized 2035 scenario with a 100\% renewable share with Static Line Rating and Dynamic Line Rating. The spatial distribution of capacities in the SLR and DLR model can be found in Fig.~\ref{fig:capacity-map-2035}}
    \centering
    \begin{tabular}{lrrrrc}
        & Plan & SLR & DLR & $\Delta$ & Unit  \\
        \hline
        Offshore Wind & 40 & 17.8 & 32.7 & 14.9 & GW \\
        Onshore Wind & 157 & 215.3 & 176.2 & --39 & GW \\
        Solar & 309 & 244.1 & 220.2 & --23.9 & GW \\
        Battery Discharge & - & 23.7 & 20.5 & --3.2 & GW \\
        Battery Storage & - & 176.6 & 237.6 & 61.1 & GWh \\
        Hydrogen Electrolysis & - & 75.7 & 61.5 & --14.2 & GW \\
        Hydrogen Fuel Cell & - & 77.4 & 74.8 & --2.6 & GW \\
        Hydrogen Storage & - & 15.5 & 13.1 & --2.4 & TWh \\
        \hline
    \end{tabular}
    \label{tab:installed capacity_2035}
\end{table}
In the following, we take a closer look at the fully decarbonized power system scenario. Table~\ref{tab:installed capacity_2035} shows the total capacities for the DLR and the SLR model at 100\% renewable share.
In addition to Fig.~\ref{fig:capacity-2035}, Table~\ref{tab:installed capacity_2035} puts the total capacities into perspective and compares them to the expansion plans of the German government. It shows that both the DLR and the SLR model tend to build less offshore wind and solar and more onshore wind capacities than planned by the German government.

\begin{figure}[!h]
    \centering
    \includegraphics[width=0.7\linewidth]{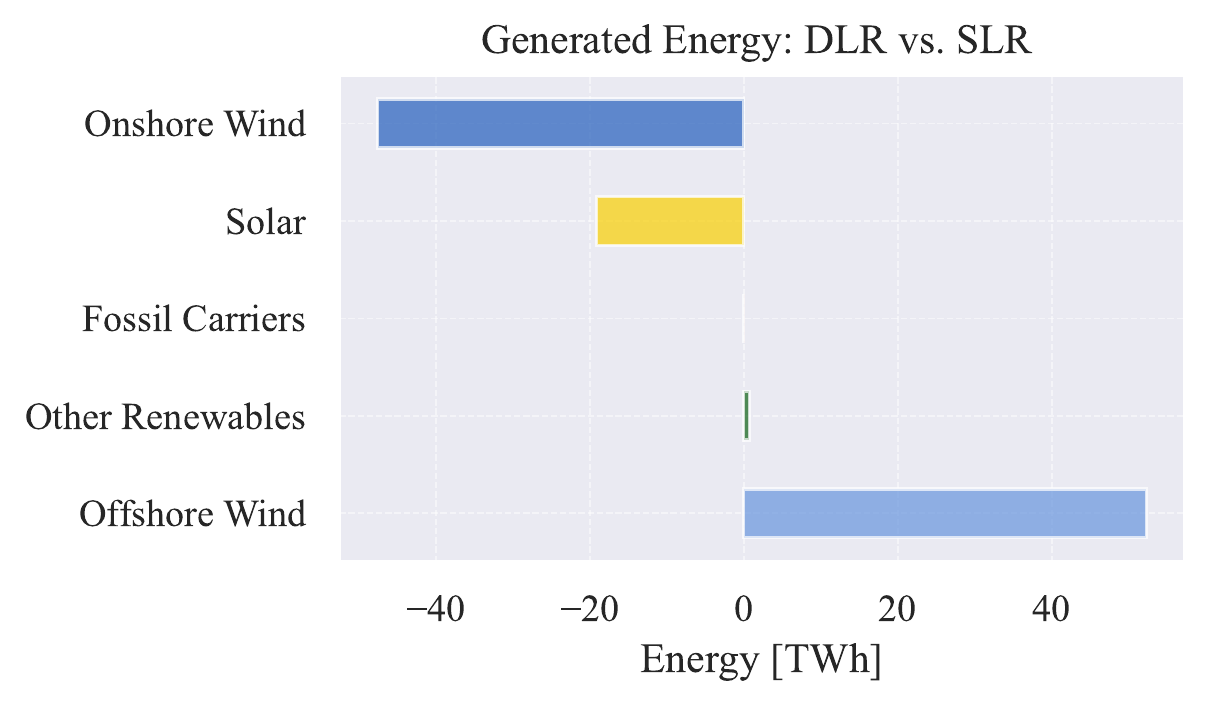}
    \caption{The figure shows the total energy difference when going from SLR to DLR. Negative values mean less generation in the DLR model.}
    \label{fig:RE1-operation-difference}
\end{figure}

In addition, the total electricity generation difference between DLR and SLR is illustrated in Fig.~\ref{fig:RE1-operation-difference}.
We see how additional offshore wind generation in the DLR model substitutes solar and onshore wind generation. Despite less renewable capacity in the DLR model (see Table~\ref{tab:installed capacity_2035}), better integration and higher capacity factors of wind power allow for a stable energy supply.
In total the DLR model saves 14~TWh in energy production due to reduced storage utilization and associated conversion losses.
This shows the advantage of investments in DLR over investments in storage infrastructure in power system encountering grid bottlenecks.
To sum up, DLR enhances the grid such that the power system can better integrate offshore wind while requiring less power generation and storage backup capacities.

\section{Limitations}
\label{sec:limitations}
The presented power system model reveals a high spatial resolution as well as detailed information about installed capacities and renewable potentials. However, it is not able to represent all the power system dynamics.

The model is isolated from other countries, neglecting cross-border exchanges.
In particular, missing cross-border lines lead to an overestimated storage infrastructure capacity expansion and curtailment.
Particularly high curtailment can be seen in the north-west of Germany at the substations ``Diele \& Dörpen West'', where much of the curtailed wind power could be exported to the Netherlands if considering cross-borer flows.
Another limitation is the modelling of the conventional fleet. We do not consider ramping constraints for coal and gas power plants. Therefore, in our model the operation of coal and gas power plants is more flexible compared to reality.
Furthermore, the optimization of the model is kept linear, neglecting power transmission losses as well as complex power flow dynamics.
Note that the losses rise quadratically with current and DLR leads to high line loading, and therefore higher transmission losses.
For this reason, we run a nonlinear power flow calculation based on the linearly optimized generation dispatch to compare transmission losses of SLR and DLR.
We calculate the losses for the models with 100\% renewable share assuming they have the highest losses.

The losses in the SLR model account for 1.7\% of the total transmission while for the DLR model they account for 3.6\%.
In absolute numbers the losses represent 166 and 60~TWh for the DLR and SLR model, respectively.
When looking at the complex power flow dynamics, the nonlinear power flow calculation converged for 99.99\% of all time steps in all regarded scenarios.

\section{Conclusion}
\label{sec:conclusion}
In this paper, we evaluate the effect of DLR on multiple study scenarios. In the first setup, we consider the German power system in a historical scenario for the year 2019 to assess the effect of DLR on the operation of the existing grid. In the second scenario, we model the year 2030 enforcing an 80\% renewable share throughout the simulation year in order to assess the effect of DLR on optimal infrastructure planning. In the third scenario, we extend the second scenario and model the transition from 2030--2035 by gradually varying the renewable share from 80--100\%. We show that DLR offers large potentials for system cost savings as well as considerable benefits for transmission system operation.

In the 2019 scenario, we observe operational cost savings of 583~m\euro{}/yr at a gas price of 27~\euro{}/MWh through DLR due to better wind power integration and a shift away from fossil energy carriers. The 2030 study case with 80\% renewable share shows that an implementation of DLR together with an optimal deployment of renewable generation saves 1.15~bn\euro{} of annual capital and operational system costs. Compared to the SLR model, the DLR model reduces the need for battery storage, which must be widely deployed, by one third. This reduction is mainly due to a better integration and exploitation of wind power and less transmission congestion. These findings are robust against network security constraints, clipping the maximum power transmission on lines to 130--200\% of the nominal capacity. Already with a rather conservative DLR capacity limit of 130\%, system cost savings sum up to 1~bn\euro{}/yr. The benefits increase further when raising the line capacity limit of DLR.

With increasing share in renewable production from 80--100\%, we see that the DLR model gradually deploys more offshore wind as well as less onshore wind and solar capacities than the SLR model. At  100\% renewable share, the implementation of DLR clearly shifts the optimal production from onshore wind and solar generation to offshore wind power leading to steady power feed-ins and therefore less storage requirements (2.2 TWh). By reducing the operation of storage technologies and their associated losses, the system saves around 14~TWh in energy production and 3.9bn\euro{}/yr in total cost. In that regard, the estimated costs of a system-wide implementation of DLR of roughly 80m\euro{}/yr are negligible, assuming DLR cost of 80k\euro{} per line km, 8\% interest rate, 30 years lifetime.

We conclude that given the urgent need for decarbonizing the German power system, DLR is a viable complement to transmission capacity expansion, not only increasing the total welfare through a non-invasive measure but also reducing grid congestion.

\section*{Acknowledgement}
We thank Tom Brown for suggesting the project and Rena Kuwahata from Ampacimon for fruitful discussions.
This paper was conducted partly for the CoNDyNet2 project, which was supported by the German Federal Ministry of Education and Research under grant number 03EK3055E.
Furthermore, we thank Breakthrough Energy for partially funding this work in the project “Hydrogen
Integration and Carbon Management in Energy System Models”.
\newpage

\appendix

\section{}
\setcounter{figure}{0}
\setcounter{table}{0}

\subsection{Study Cases Parameters}\label{tab:cases_parameters}
\begin{tabularx}{\linewidth}{>{\bfseries}l  X  X}
    \hline
    { }                      & 2019                         & 2030 - 2035                                                   \\
    \hline
    Generator capacity                                                                                                      \\expansion    &            no &                           yes \\
    CO2 Limit                & 222 mil tons                 & -                                                             \\
    CO2 Price                & -                            & 120~\euro{}/t                                                 \\
    Gas price& 27\euro{}/MWh & 27\euro{}/MWh\\
    Base load of nuclear                                                                                                    \\and lignite &           yes &                            no \\
    Renewable generation                                                                                                    \\constraint &            no &      80\% -- 100\% renewable production share \\
    Generator infrastructure & historic capacities for 2019 & capacities of 2022 with decommissioning dates later then 2030 \\
    Electrical load          & 605 TWh                      & 658 TWh                                                       \\
    \hline
\end{tabularx}

\subsection{DLR Factor}\label{dlr_factor}
In the following, we illustrate why the hourly averaged wind speed overestimates the DLR transmission capacity compared to higher resolved data. According to the IEEE standard~\cite{ieeeIEEEStandardCalculating2012},
\begin{align*}
    P_{max}\varpropto I_{max} \underset{\sim}{\varpropto} \qquad \stackrel{\varpropto}{\sim}
    \sqrt{\overline{v}_{wind}}
\end{align*}
, where $\overline{v}_{wind}$ denotes the averaged hourly wind speed.
In the following equation,
\begin{align*}
    \sqrt{\overline{v}_{wind}}=\sqrt{\frac{1}{n}\sum_{i}^{n}{v_{wind,i}}} \ge \frac{1}{n}\sum_{i}^{n}{\sqrt{v_{wind_i}}}
\end{align*}
$v_{wind,i}$ corresponds to an sub-hourly wind speed data point going from $i$ to $n$, i.e. from $1$ to $6$ when regarding 10-minute wind speed data. The Equation shows that the root of the averaged sub-hourly wind speed is greater equal the average of the rooted sub-hourly wind speed.

\subsection{Scenario Plots}
\label{sec:scenario-plots}

\begin{figure}[!h]
    \centering
    \includegraphics[width=.7\linewidth]{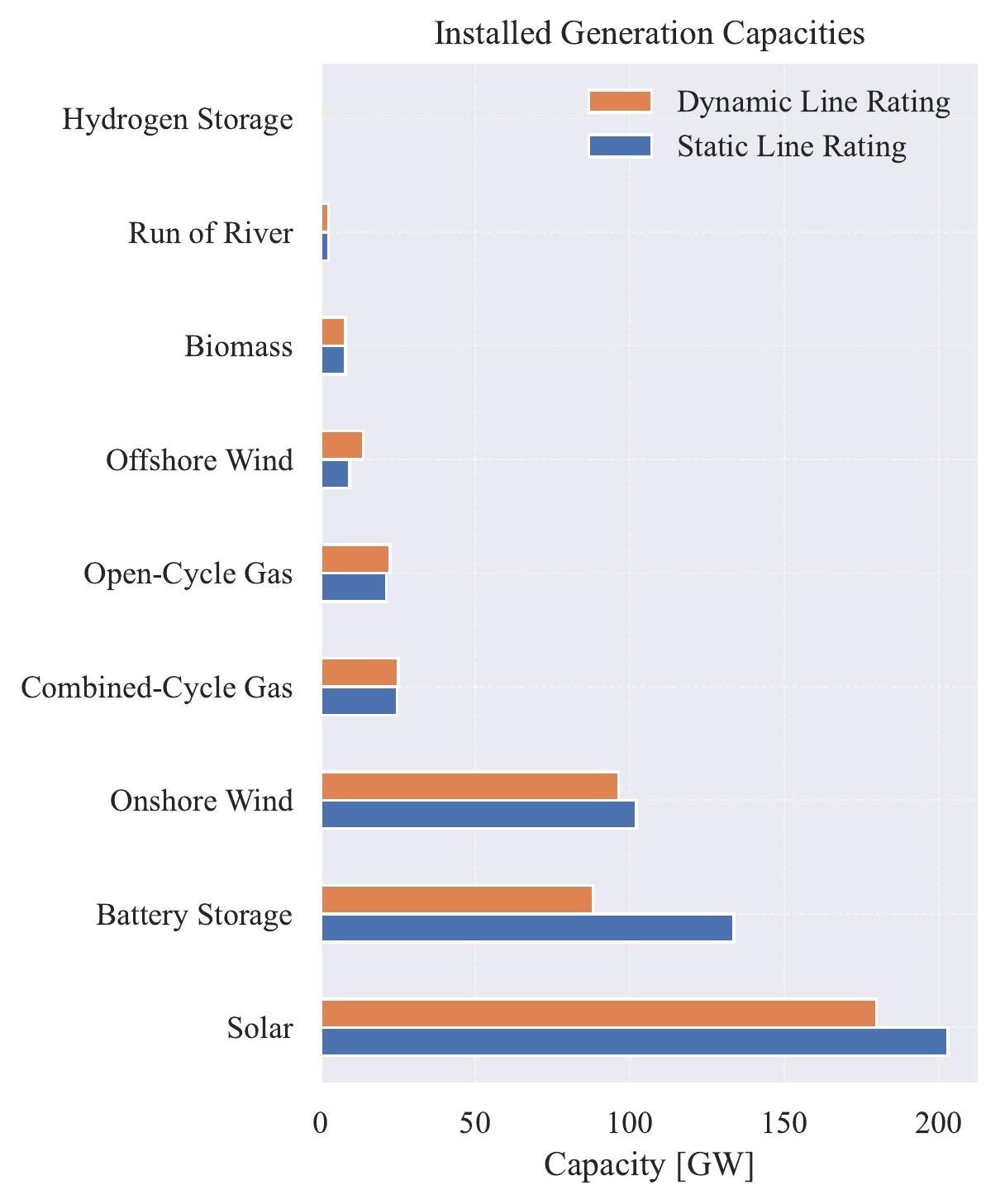}
    \caption{Optimized capacities for the 2030 scenario with 80\% renewable share.}
    \label{fig:capacity-bar-2030}
\end{figure}

\begin{figure}[!h]
    \centering
    \includegraphics[width=\linewidth]{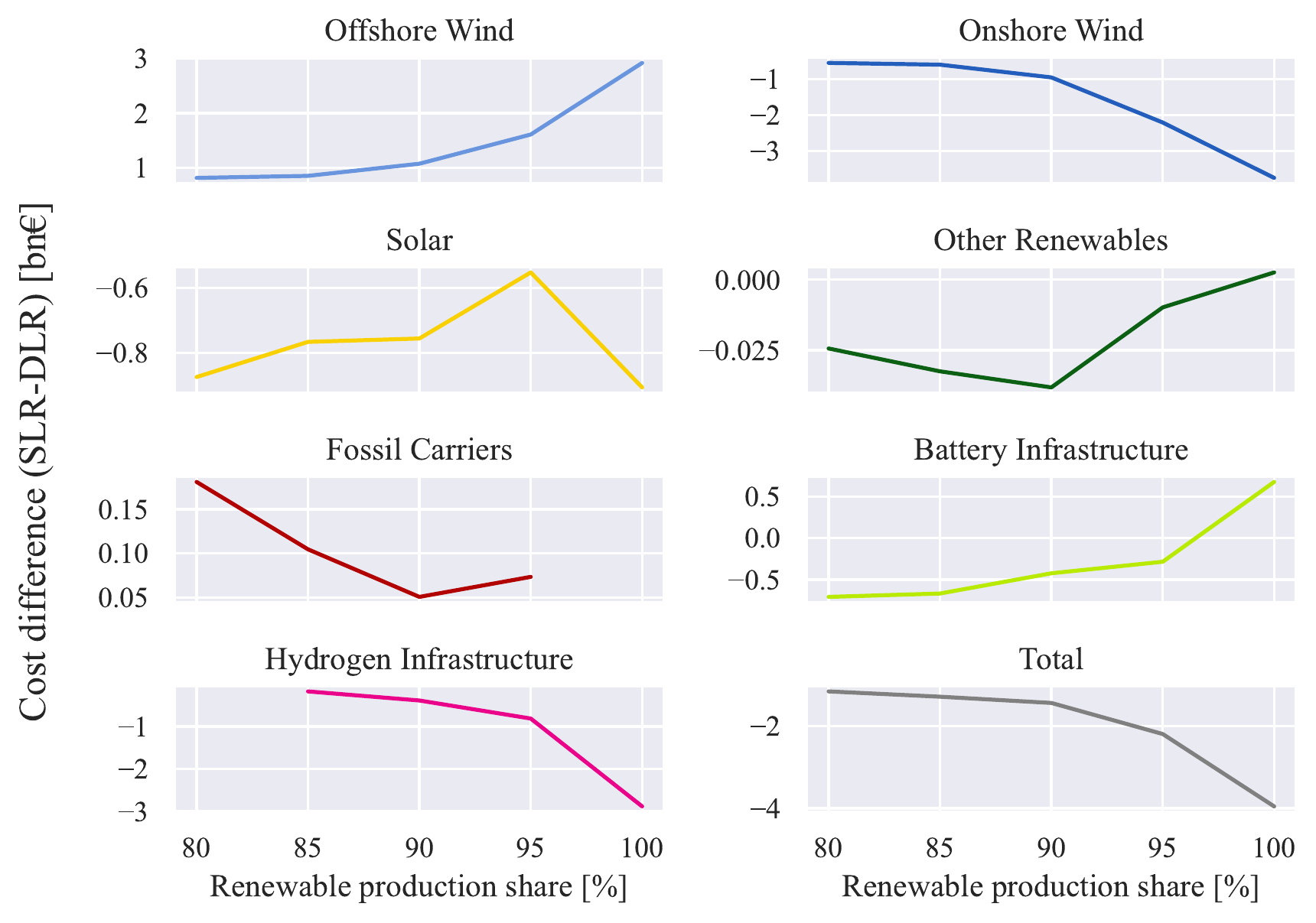}
    \caption{Difference in capital and operational expenditures when going from SLR to DLR for different renewable shares. Positive costs represent higher expenses of DLR.}
    \label{fig:cost-sensitivity-2035}
\end{figure}

\begin{figure}[!h]
    \centering
    \includegraphics[width=\linewidth]{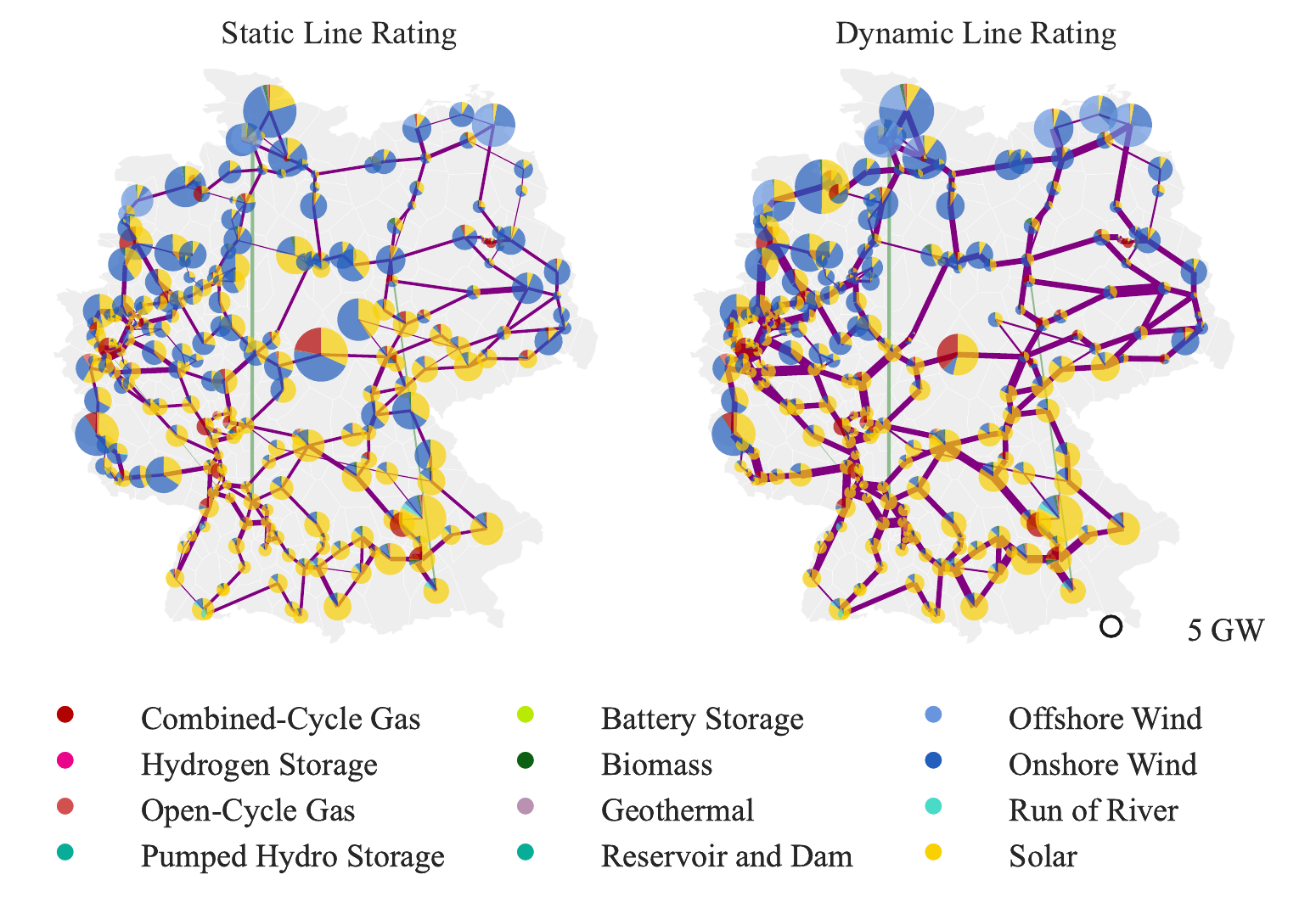}
    \caption{Optimized capacities for the 2035 scenario with 100\% renewable share.}
    \label{fig:capacity-map-2035}
\end{figure}

\clearpage
\bibliographystyle{elsarticle-harv}
\bibliography{main.bib}

\end{document}